\documentclass[
  aps,
  prl,
  reprint,
  amsmath,
  amssymb
]{revtex4-1}
\setcitestyle{square}
\usepackage[pdftex]{graphicx}
\usepackage[pdftex,bookmarks,colorlinks,citecolor=blue,linkcolor=blue]{hyperref}
\usepackage[pdftex,svgnames,usenames,dvipsnames]{xcolor}
\usepackage[utf8]{inputenc}
\usepackage[T1]{fontenc}
\usepackage{siunitx}
\usepackage{url}
\usepackage{bm}
\usepackage[normalem]{ulem} 

\newcommand\figfolder{.}

\begin{document}

\title{Force transmission and the order parameter of shear thickening}

\author{Romain Mari}
\affiliation{Univ. Grenoble-Alpes, CNRS, LIPhy, F-38000 Grenoble, France}
\author{Ryohei Seto}
\affiliation{Department of Chemical Engineering, Kyoto University, Kyoto, Japan}

\date{\today}


\begin{abstract}
The origin of the abrupt shear thickening observed in some dense suspensions has been recently argued to be a transition from 
frictionless (lubricated) to frictional interactions between immersed particles. The Wyart-Cates rheological model, built on this scenario, 
introduced the concept of fraction of frictional contacts \(f\) as the relevant 
order parameter for the shear thickening transition.
Central to the model is the ``equation-of-state'' relating \(f\) to the applied stress \(\sigma\), 
which is directly linked to the distribution of the normal components of non-hydrodynamics interparticle forces.
Here, we develop a model for this force distribution, based on the so-called \(q\)-model that we borrow from granular physics.
This model explains the known \(f(\sigma)\) in the simple case of sphere contacts displaying only sliding friction, but also predicts 
strong deviation from this ``usual'' form when stronger kinds of constraints are applied on relative motion.
We verify these predictions in the case of contacts with rolling friction, in particular a broadening of the stress range 
over which shear thickening occurs.
We finally discuss how a similar approach can be followed to predict \(f(\sigma)\) in systems with other variations 
from the canonical system of monodisperse spheres with sliding friction, in particular the case of large bidispersity.
\end{abstract}
\maketitle


\section{Introduction}

Shear thickening is an increase of the viscosity with applied stress observed in the flow of some dense suspensions of hard particles, 
which size lies usually in the tens of nanometer to tens of micrometer range~\cite{Barnes_1989,Brown_2014}. 
The viscosity \(\eta \) increase can be arbitrarily small, at low volume fractions \(\phi \) 
(typically when the particles occupy less than half of the total volume of the suspensions), 
to arbitrarily large, at high volume fractions.
As a function of the shear rate \(\dot\gamma \), the viscosity shows two distinct behaviors: 
continuous shear thickening (CST), i.e., \(\mathrm{d}\eta/\mathrm{d}\dot\gamma<\infty \), 
for volume fractions below a critical \(\phi_\mathrm{c}\), 
and discontinuous shear thickening (DST), i.e., a jump of the viscosity at a given shear rate, for \(\phi > \phi_\mathrm{c}\).

In the past few years, advances in the understanding of the microscopic physics of 
dense suspensions of hard particles led to the development of the frictional transition scenario 
to explain shear thickening~\cite{morris_2018}.
In this scenario, shear thickening appears when two kinds of interparticles forces are present: 
a repulsive force (stemming from coated polymer brushes, electrical double layer, etc.)
and dry-like frictional forces, 
usually thought as a consequence of direct contact between particles 
by rupture of the lubrication film~\cite{fernandez_microscopic_2013,seto_discontinuous_2013,heussinger_shear_2013,lin_hydrodynamic_2015,royer_rheological_2016,clavaud_revealing_2017,comtet_pairwise_2017}.

The general picture is as follows. 
At low applied stresses, the repulsive forces maintain particles separated by a finite gap, 
and in consequence particles interact via the repulsive force, which is a normal force, and via lubrication, which is a viscous 
force with normal and tangential components. 
In such a system the viscosity only diverges at the volume fraction of jamming for frictionless particles. 
On the other hand, at high applied stresses,
the repulsive forces are usually overcome by confining forces from surrounding particles in shear
and contacts proliferate, making the particles interact 
via a static force  with both normal and tangential components.
The viscosity then diverges at a volume fraction below the frictionless jamming transition, 
a feature typical of systems with forces restricting particles rotational degrees of freedom~\cite{maxwell_calculation_1864,silbert_geometry_2002,gallier_rheology_2014,Mari_2014}.

While initial suggestions for this scenario came from numerical simulations, a theoretical description 
by Wyart and Cates~\cite{Wyart_2014} further enlightened the exact relation between CST and DST.
The Wyart-Cates (WC) theory relies on a scalar constitutive model for shear-thickening suspensions, 
relating the steady-state viscosity to the shear stress.
(Nakanishi et al.~\cite{Nakanishi_2012} also introduced a similar type of constitutive model,
which was used to solve fluid dynamics problems of shear-thickening suspensions.)
In this model, the viscosity of the suspension diverges algebraically,
\(\eta  \sim  {\big(\phi_\mathrm{J}(f)-\phi\big)}^{-\alpha}\),
by approaching the jamming volume fraction \(\phi_\mathrm{J}\), 
as is often proposed close to the jamming transition~\cite{krieger_mechanism_1959,Zarraga_2000,Boyer_2011,Lerner_2012}. 
The value of the exponent \(\alpha \) is debated in the literature~\cite{Lerner_2012,gallier_fictitious_2014,Mari_2014,ness_flow_2015}, 
but this is not the focus of the present discussion; 
the most typical value in the literature is \(\alpha=2\)~\cite{ness_flow_2015,hermes_unsteady_2016,guy_constraint-based_2018}. 
The specificity of the WC theory is that \(\phi_\mathrm{J}\) depends on a microscopic order parameter \(f\), 
the so-called fraction of frictional contacts, that is, the proportion of nearest neighbor interactions which are frictional, 
as opposed to lubricated.
The jamming volume fraction is linearly moving with \(f\) as \(\phi_\mathrm{J}(f) = (1-f)\phi^0_\mathrm{J} + f\phi^1_\mathrm{J}\), 
in between two limiting values, \(\phi^0_\mathrm{J}\) for only lubricated interactions 
and \(\phi^1_\mathrm{J}<\phi^0_\mathrm{J}\) when only frictional contacts are present.
Finally, the competition between confining forces due to shear and repulsive forces at the microscopic level implies that \(f\) 
is a sigmoidal-like function of the shear stress \(\sigma \), 
increasing from \(0\) at vanishing stress to \(1\) in the large stress limit.
In practice, it has been 
numerically observed that 
\begin{equation}
    f(\sigma) \approx \exp\left(-c F_\mathrm{c} a^2/\sigma\right)
    \label{eq:exp_fofsigma}
\end{equation}
with \(F_\mathrm{c}\) the repulsive force at contact, \(a\) the typical particle radius, 
and \(c\) some constant of order unity~\cite{ness_shear_2016,singh_constitutive_2018}.
(Note that this differs from the initial choice made by Wyart and Cates themselves, who picked 
\(f(p)= 1-\exp(-p)\), with \(p\) the particle pressure, as giving a representative rheological behavior~\cite{Wyart_2014}.)
Remarkably, the \(\phi \) dependence is observed to be very weak, 
at least over the range of \(\phi \) where shear thickening occurs~\cite{Mari_2014}.
It has been however argued that this form should be modified to \(f(\sigma) \approx f_\mathrm{max}\exp(-c F_\mathrm{c} a^2/\sigma)\) 
with \(0 < f_\mathrm{max} \leq 1\) to get a better fit of the WC model to experimental flow curves 
at low volume fractions~\cite{royer_rheological_2016}, 
or \(f(\sigma) \approx \exp[-c {(F_\mathrm{c} a^2/\sigma)}^\beta]\) with \(\beta<1\) 
to account for a broader stress range for thickening~\cite{guy_towards_2015}.
Altogether, the relations \(\eta(\phi, \phi_\mathrm{J})\), \(\phi_\mathrm{J}(f)\) and \(f(\sigma)\) 
form the scalar WC constitutive model, 
which exhibits a shear-thickening rheology with the major experimentally observed features. 
In particular, it allows the difference in viscosity between the unthickened (\(f\approx 0\)) 
and thickened (\(f\approx 1\)) to be arbitrarily large 
provided one increases the volume fraction close enough to \(\phi^1_\mathrm{J}\).
Together with~Eq.~\ref{eq:exp_fofsigma}, this implies  
that there is a volume fraction \(\phi_\mathrm{c}<\phi^1_\mathrm{J}\) above which there is a stress window 
such that \(\mathrm{d}\eta/\mathrm{d}\sigma > \eta/\sigma \), that is, DST, 
while for \(\phi<\phi_\mathrm{c}\) only CST is observed~\cite{Wyart_2014}.

At a more quantitative level, early experimental tests of the WC model showed it is quite succesful 
at fitting actual flow curves of model shear-thickening suspensions~\cite{guy_towards_2015,royer_rheological_2016,hermes_unsteady_2016,ness_shear_2016}, 
made of spherical, reasonably monodisperse particles. 
A tensorial extension of the WC model has also been proposed 
and validated against particle-based simulations~\cite{singh_constitutive_2018}.
Here again, the simulations were using moderate polydispersity 
(bidispersity with a size ratio of 1.4), 
and only sliding friction was considered, in order to model spherical particles with only moderate surface roughness.
Very recently, however, Guy and coworkers showed that the WC model, while qualitatively still correct, 
is a poor quantitative model for highly polydisperse suspensions~\cite{guy_testing_2019}, 
and part of their diagnosis points 
to a failure of Eq.~\ref{eq:exp_fofsigma} when the suspension is far from being monodisperse.

Unsurprisingly, the WC model is quite sensitive to the functional form for \(f(\sigma)\).
It then matters to understand in what cases  Eq.~\ref{eq:exp_fofsigma} fails and why it does.
In this work, we show how \(f(\sigma)\) is related to the nature of particles contacts at the microscopic scale.
We base our analysis on the relation between the distribution of 
non-hydrodynamic interparticle forces  
and the fraction of frictional contacts, which exponential tail at large forces has been argued to 
be the origin of Eq.~\ref{eq:exp_fofsigma}~\cite{guy_towards_2015,royer_rheological_2016,ness_shear_2016}. 
We borrow, and adapt to shear-thickening suspensions, the celebrated \(q\)-model introduced in the context 
of the statistical description 
of forces chains in sandpiles~\cite{liu_force_1995,coppersmith_model_1996}.
We show that, at large forces, the force distribution blatantly departs from exponential decay during thickening, 
if thickening occurs between two states differing strongly in the heterogeneity of their force propagation.
This, in turn, implies that for such system \(f(\sigma)\) goes from \(0\) to \(1\) over a much broader stress range, 
and is better fitted by a stretched exponential \(f(\sigma) \approx \exp[-c {(F_\mathrm{c} a^2/\sigma)}^\beta]\) 
with \(\beta<1\).
We then test this prediction with particle-based simulations of suspensions of particles interacting with rolling 
as well as sliding friction in the shear-thickened state, 
and show that our model captures qualitatively the effect 
of a large rolling friction coefficient.
This is important for the many systems for which contact with simple sliding friction is only a rough approximation. 
In particular, for many real non-model suspensions processed in industry, effective rolling friction may be at work, 
for instance for non-spherical particles (e.g.\ cement particles in fresh concrete~\cite{Lootens_2004}, 
sugar grains in molten chocolate~\cite{blanco_conching_2019}).
We finally discuss predictions of modified \(q\)-models for highly-polydisperse suspensions as the ones of Guy et al.~\cite{guy_testing_2019}.


\section{Simulation model}

In this work, we use the so-called Critical Load Model (CLM), 
which is the simplest model exhibiting CST and DST~\cite{Mari_2014}.
We simulate systems of 500 bidisperse particles (radii \(a\) and \(1.4a\) in equal
volume proportions), sheared in a tri-periodic cubic box with the Lees-Edwards boundary conditions 
under constant shear stress \(\bar\sigma \)~\cite{mari_nonmonotonic_2015,Seto_2019}.
Particles are subject to a Stokes drag and interact through short-range pairwise 
hydrodynamic forces (lubrication) and frictional contact forces.

Expressions for the lubrication forces are given in~\cite{Mari_2014}.
The force on a particle \(i\) with radius \(a_i\) 
from a contact with particle \(j\)
can be decomposed in normal and tangential components, 
coming respectively from the hard core interaction and sliding friction, 
while the torque has contributions from sliding and rolling friction
\begin{equation}
  \label{eq:contact_model}
  \begin{split}
  \bm{F}_{\mathrm{C}}^{(i,j)} & = \bm{F}_{\mathrm{C,nor}}^{(i,j)} + \bm{F}_{\mathrm{C, tan}}^{(i,j)} \\
  \bm{T}_{\mathrm{C}}^{(i,j)} & = a_i \bm{n}_{ij} \times \big(\bm{F}_{\mathrm{C,tan}}^{(i,j)}  + \bm{F}_{\mathrm{C,roll}}^{(i,j)}\big).
  \end{split}
\end{equation}
Contacts fulfill Coulomb's friction laws 
\(\bigl|\bm{F}_{\mathrm{C,tan}}^{(i,j)} \bigr| \leq \mu_\mathrm{s} \Delta F_{\mathrm{C,nor}}^{(i,j)}\) and 
\(\bigl|\bm{F}_{\mathrm{C,roll}}^{(i,j)} \bigr| \leq \mu_\mathrm{r} \Delta F_{\mathrm{C,nor}}^{(i,j)}\) with 
sliding (resp.\ rolling) friction coefficient \( \mu_\mathrm{s} \) (resp.\ \(\mu_\mathrm{r}\))
and \(\Delta F_{\mathrm{C,nor}}^{(i,j)} = \max(0, \bigl|\bm{F}_{\mathrm{C,nor}}^{(i,j)}\bigr| - F_\mathrm{c})\).
Here, \(F_\mathrm{c}\) is a critical load below which contacts are frictionless, which is giving this model a shear-thickening rheology, 
with an onset stress \(\sigma_\mathrm{c} \sim  F_\mathrm{c}/a^2\)~\cite{Mari_2014}.
Note that \(\bm{F}_{\mathrm{C,roll}}^{(i,j)}\) is a quasi-force,
which generates only the rolling torque.
Finally, the force components \(\bm{F}_{\mathrm{C,nor}}^{(i,j)}\), \(\bm{F}_{\mathrm{C, tan}}^{(i,j)}\) 
and \(\bm{F}_{\mathrm{C,roll}}^{(i,j)}\) are obtained with virtual springs in a Cundall-Strack manner~\cite{Cundall_1979}, 
following the algorithm of~\cite{Luding_2008}.

\section{Fraction of frictional contacts and force distribution}

The fraction of frictional contacts can be tied to the probability distribution 
\(P(F)\) of non-hydrodynamic normal forces between particles in suspensions. 
As in a shear-thickening suspension, there is a repulsive force imposing a minimal load \(F_\mathrm{c}\) 
on a pair of particles for them to make a contact, 
\(f\) is nothing but the proportion of non-hydrodynamic normal forces exceeding \(F_\mathrm{c}\), 
that is, \(f = \int_{F_\mathrm{c}}^\infty \mathrm{d}F P(F)\).
Changing variable in the integral to the dimensionless \(\tilde{F} = F/\sigma a^2\), this becomes 
\begin{equation}
    f = \int_{F_\mathrm{c}/\sigma a^2}^\infty \mathrm{d}\tilde{F} \tilde{P}(\tilde{F}),
    \label{eq:f_and_PF}
\end{equation}
with \(\tilde{P}\) the distribution for \(\tilde{F}\).

By differentiating Eqs.~\ref{eq:exp_fofsigma} and~\ref{eq:f_and_PF} with respect to \(\sigma \), 
one obtains that the force distribution must be exponential to ensure Eq.~\ref{eq:exp_fofsigma}:
\begin{equation}
    \tilde{P}(\tilde{F}) = c \exp\bigl(-c\tilde{F}\bigr).
    \label{eq:PofF_exp}
\end{equation}
The force distribution in a flowing suspension of micrometer grains is thus far unaccessible to experiments.
(It is possible to measure it for dry assemblies of millimeter or larger particles 
under static~\cite{liu_force_1995,mueth_force_1998,lovoll_force_1999,blair_force_2001,erikson_force_2002,corwin_structural_2005} 
or quasi-static~\cite{howell_stress_1999} conditions.)
However, it is easily measured in numerical simulations, and recent works suggested that indeed, 
in a shear-thickening suspension of particles with sliding friction only,
the force distribution under flow 
has an exponential decay, however only seen at large forces, with a full distribution argued 
to be well fitted by \(\tilde{P}(\tilde{F}) = a\big(1-b\exp(-\tilde{F}^2)\big)\exp(-c\tilde{F})\)~\cite{ness_shear_2016,radhakrishnan_force_2019}.

\begin{figure}[t]
    \centering
    \includegraphics[width=0.42\textwidth]{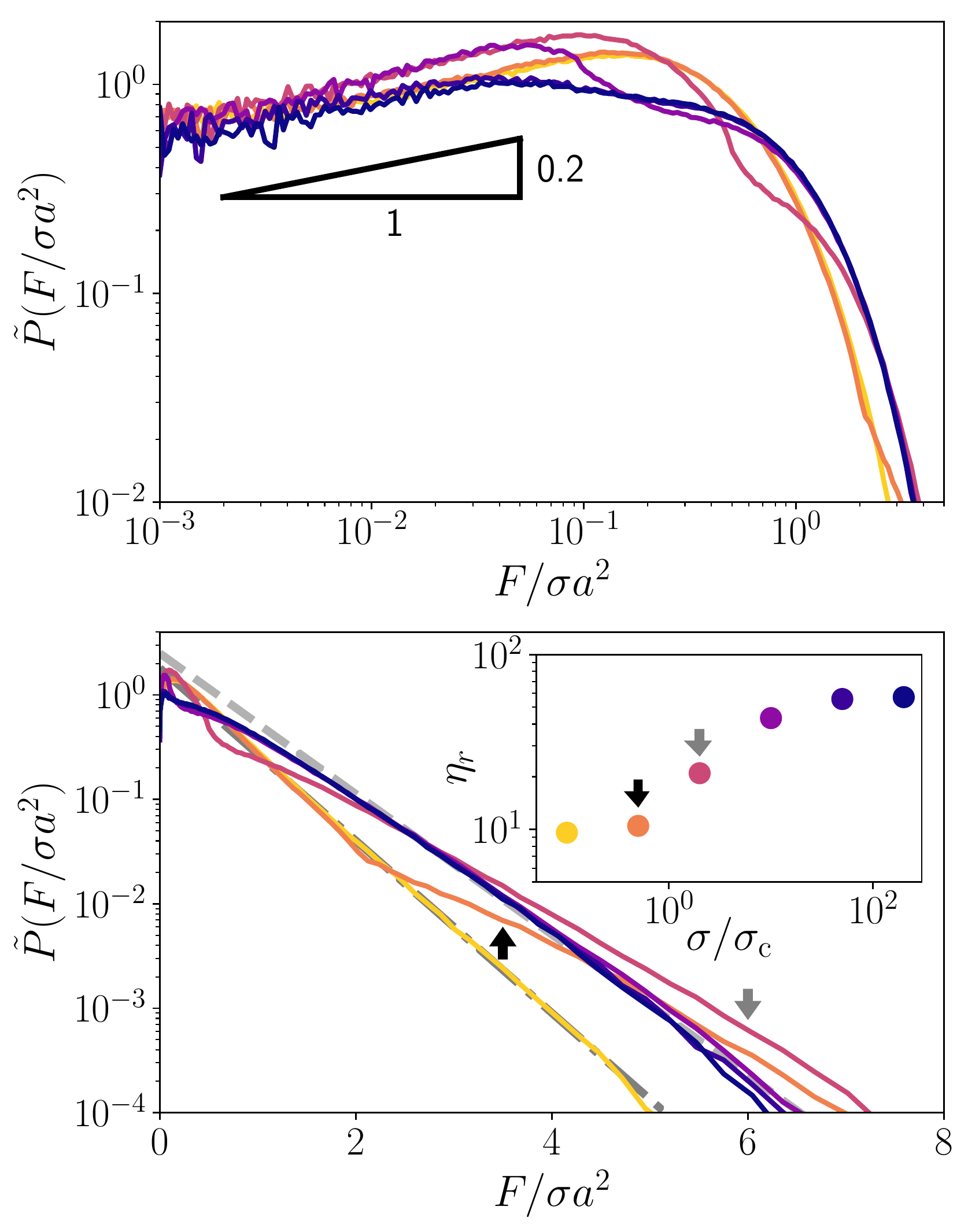}
    \caption{Force distribution \(\tilde{P}(F/\sigma a^2)\) in simulations of the CLM model with sliding friction coefficient \(\mu_\mathrm{s} = 1\) but without rolling friction (rolling friction coefficient \(\mu_\mathrm{r}=0\)), 
    in log-log (\textbf{top}) and log-lin (\textbf{bottom}),
    for several values of the applied stress \(\sigma/\sigma_\mathrm{c}\), crossing the thickening transition.
    The color associated to each stress corresponds to the one given in the inset.
    In the inset, the viscosity as a function of the stress, for the same conditions than in the main plots.}\label{fig:PofF_simus_sliding}
 \end{figure}
In Fig.~\ref{fig:PofF_simus_sliding}, we show the distribution \(\tilde{P}(\tilde{F})\) 
of normal contact forces for our CLM simulations 
with \(\mu_\mathrm{s} = 1\) and \(\mu_\mathrm{r}=0\) (no rolling friction). 
At low forces, there is a maximum below which the distribution rather follows 
a power-law behavior \(\tilde{P}(\tilde{F})\sim \tilde{F}^\theta \) 
(which is a common feature observed from dry granular packings~\cite{liu_force_1995} 
to flowing of frictionless suspensions~\cite{lerner_toward_2012}).
The exponent is small \(\theta\approx 0.2\) and does not seem to depend on the applied stress.

On the contrary, at large forces it is apparent that the force distribution 
depends on stress. 
While both in the unthickened and thickened states (resp.\ yellow and dark purple curves in Fig.~\ref{fig:PofF_simus_sliding}), 
the large force tail of the distribution is exponential 
\(\tilde{P}(\tilde{F}) \sim \exp(-c\tilde{F})\), 
the decay constant \(c\) is smaller in the thickened (\(c_0\approx 1.5\), dashed line) 
than in the unthickened state (\(c_1\approx 2\), dotted dashed line).
Moreover, a closer look at the force distributions at intermediate values of the applied stress 
(for instance \(\sigma/\sigma_\mathrm{c} = 0.5\), indicated by the black arrow) 
reveals that they actually do not have a simple exponential decay at large forces, 
but rather interpolate between the frictionless distribution with decay constant \(c_0\) at low forces  
and the frictional one with decay constant \(c_1\) at large forces.
For some values of stress around \(\sigma/\sigma_\mathrm{c} = 2\), however, 
\(\tilde{P}(\tilde{F})\) has an even slower decay at large forces, with \(c<c_1\) (indicated by the grey arrow).

\begin{figure}[t]
  \centering
  \includegraphics[width=0.42\textwidth]{\figfolder/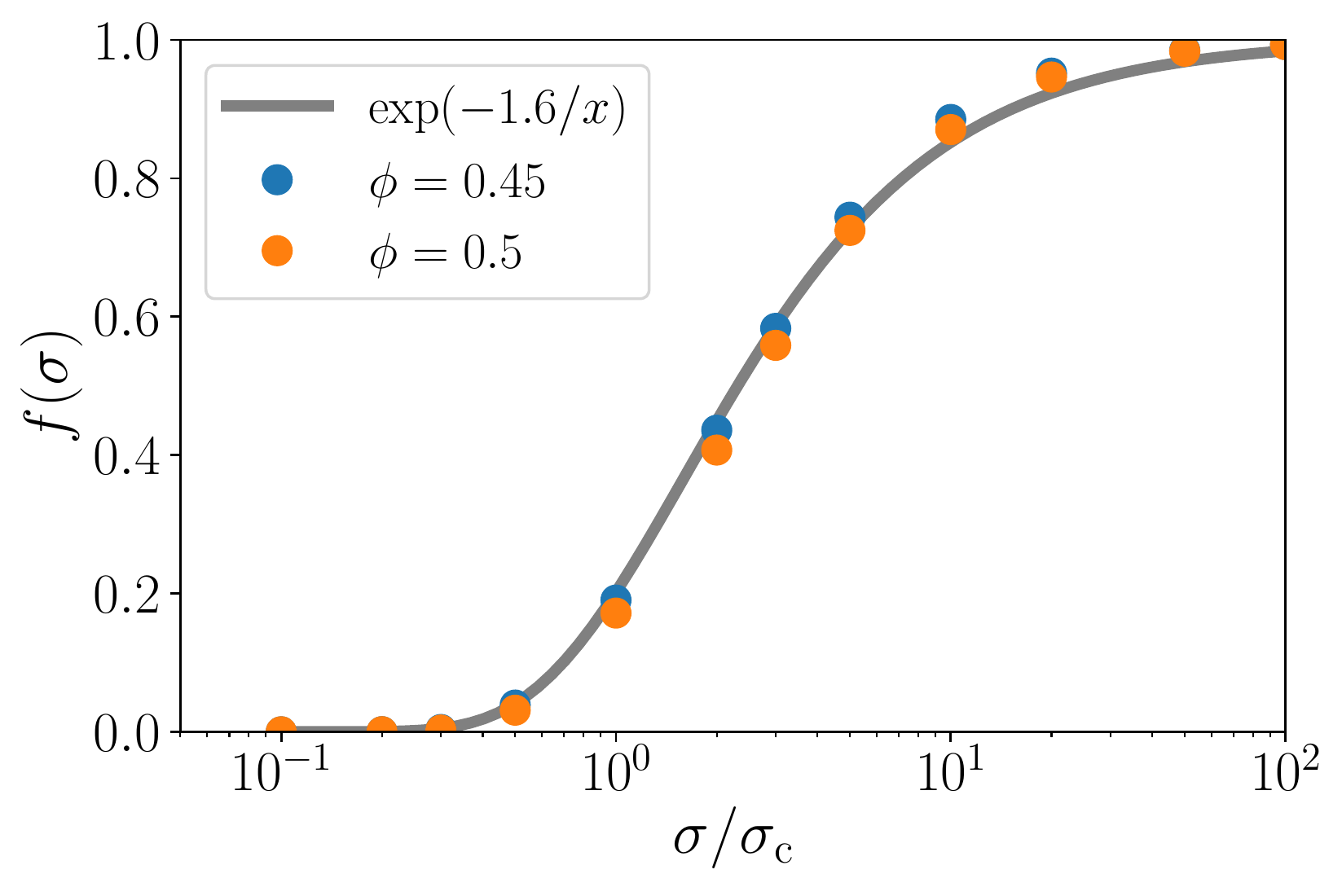}
  \caption{Fraction of frictional contacts as a function of the applied stress for the critical load model 
    with sliding friction coefficient \(\mu_\mathrm{s}=1\), and no rolling friction, at volume fractions \(\phi=0.45\) and \(\phi=0.5\) (colored symbols).
    A fit to a form \(\exp(-\sigma^\ast/\sigma)\) is shown in dark grey solid line, with \(\sigma^\ast=1.6\).}\label{fig:fofsigma_sliding}
\end{figure}

Nonetheless, the deviation from a simple exponential form common 
to all stresses is small enough so that Eq.~\ref{eq:exp_fofsigma}, 
with a value \(c=1.6\) in between \(c_0\) and \(c_1\), 
provides a good fit to the data, as shown in Fig.~\ref{fig:fofsigma_sliding}. 
In this figure, we also confirm that \(f(\sigma)\) is quite insensitive to the volume fraction, as already shown previously~\cite{Mari_2014}.
Because Eq.~\ref{eq:exp_fofsigma} is such a good approximation to the observed \(f(\sigma)\), 
the WC model is (quite surprisingly owing to its relative simplicity) successful in 
its quantitative agreement with observed rheological data~\cite{royer_rheological_2016,ness_shear_2016,singh_constitutive_2018}.

However, we may see the glass half empty, and wonder whether there could be cases, 
different from the quite model system of monodisperse spheres with sliding friction, 
where the force distribution \(\tilde{P}(\tilde{F})\) is drifting further from the 
dry-granular-like exponential decay at large forces, 
so that \(f(\sigma)\) is far from Eq.~\ref{eq:exp_fofsigma}.
This will be the motivation for introducing, in the next section, 
a model for the non-hydrodynamic force distribution in shear-thickening suspensions.


 \section{Force distribution model}

In this section, we show that one can rationalize the force distribution in 
steady flows and its stress dependence across shear thickening,
and hence the constitutive relation between the fraction of frictional contacts 
and the applied stress, 
with a minimal extension of the seminal \(q\)-model 
of force propagation in a granular packing~\cite{liu_force_1995,coppersmith_model_1996}.

\subsection{\(q\)-model}

The \(q\)-model was initially introduced as a minimal model of grain packings 
exhibiting the normal contact force distribution 
observed experimentally for dry granular matter~\cite{coppersmith_model_1996}.
It is made of layers of \(N\) sites transmitting (scalar) forces downwards randomly through 
the bonds of a lattice, with a periodic boundary condition in the horizontal direction, i.e., \(N+1\equiv 1\).
In this work, we will only consider the \(q\)-model (and its extension introduced in the next subsection) 
on a lattice such that site \(i\) in layer \(D+1\) is connected to site \(i\) and \(i+1\) in layer \(D\).
It then receives a force \(F(i, D+1)\) 
\begin{equation}
	F(i, D+1) = a_{i,i}(D) F(i, D) + a_{i+1,i}(D) F(i+1, D),
\end{equation}
with \(a_{i,j}\)'s random positive couplings which satisfy force conservation on each site, 
that is, \(a_{i,i}(D) + a_{i,i-1}(D)=1\), \(\forall i, D\).

We consider here a coupling distribution \(\mathcal{P}(a_{i,i}, a_{i,i-1})\) such that every site transmits the same proportion \(1/2< q < 1\) 
to one of its neighbors and \((1-q)\) to the remaining one, which is
\begin{equation}
\begin{split}
  \mathcal{P}(a_{i,i}, a_{i,i-1}) = & \frac{1}{2} \big\{\delta[a_{i,i} - (1-q)] \delta[a_{i,i-1} - q]\\
   &\quad  + \delta[a_{i,i} - q] \delta[a_{i,i-1} - (1-q)]\big\}.
\label{eq:Pofq_qmodel}
\end{split}
\end{equation}
This situation was initially considered by~\citet{coppersmith_model_1996}, who found it representative 
of the behavior of the \(q\)-model, as long as the propagation is not singular 
\(q\neq 1\) or \(1/2\), i.e., transmission to a single neighbor or strictly even transmission between neighbors.

Of course, in an actual system forces applied on grains are vectors, not scalars. 
This somewhat oversimplifying nature of \(q\)-model leads to issues on the nature of force propagation at a macroscopic level 
predicted by the model~\cite{claudin_models_1998}, but it nonetheless is successful in predicting the normal force distribution 
of actual packings, at least qualitatively.
The distribution of forces \(P_D(F)\) in layer \(D\) converges for \(D\to\infty \) to \(P_\infty(F)=f^\theta \exp(-cf/\bar{F})\), 
where \(\bar{F}\) is the average force, \(c\) and \(\theta \) constants, whatever the initial distribution \(P(F)\) is in layer \(D=0\)
(provided that \(\int_0^\infty \mathrm{d}F FP(F)=\bar{F}\))~\cite{coppersmith_model_1996}.
As discussed earlier, the power law at low forces and exponential decay at large forces are well-known features 
of experimental or numerical realizations of granular packings~\cite{liu_force_1995,radjai_force_1996,mueth_force_1998,blair_force_2001}.

\subsection{Bi-\(q\)-model}

Force transmission in a flowing dense suspension shares many features with the one in dry granular packings. 
It is characterized by a local conservation law (force and torque balances on each grain) and a disorder in the contact network. 
More precisely in the case of the flowing suspension, the force and torque balances are achieved with both hydrodynamic 
and non-hydrodynamic forces (contact and surface forces), but because hydrodynamic forces are small compared to other forces
(thanks to the high volume fraction of these systems), 
force and torque balances are almost achieved by non-hydrodynamic forces alone. 
Under shear, the contact network of a thickening suspension acquires a small anisotropy~\cite{Mari_2014}, 
which is not observed in idealized packings of dry grains under isotropic compression, 
but is reminiscent of packings under gravitational load~\cite{liu_force_1995}.
There is nonetheless an important difference between the two kinds of systems: 
whereas particles in dry granular always interact with the same contact laws, 
irrespective of the local loads, particles in a shear-thickening suspension 
are interacting via frictionless contacts 
under locally small loads and frictional contacts 
under large loads~\cite{seto_discontinuous_2013,fernandez_microscopic_2013, heussinger_shear_2013, lin_hydrodynamic_2015,royer_rheological_2016,clavaud_revealing_2017,comtet_pairwise_2017}.

The force network of frictional systems is sparser, or perhaps more accurately more heterogeneous, 
compared to the one of frictionless systems, which can be seen as a remnant of the jamming transition occurring 
with less contacts per particles for frictional than for frictionless systems~\cite{silbert_geometry_2002,shundyak_force_2007}.
As we now show, this translates directly for shear-thickening suspensions, in which force transmission across the contact network 
is more homogeneous (or even) below thickening than above.

\begin{figure}[t]
  \centering
  \includegraphics[width=0.49\textwidth]{\figfolder/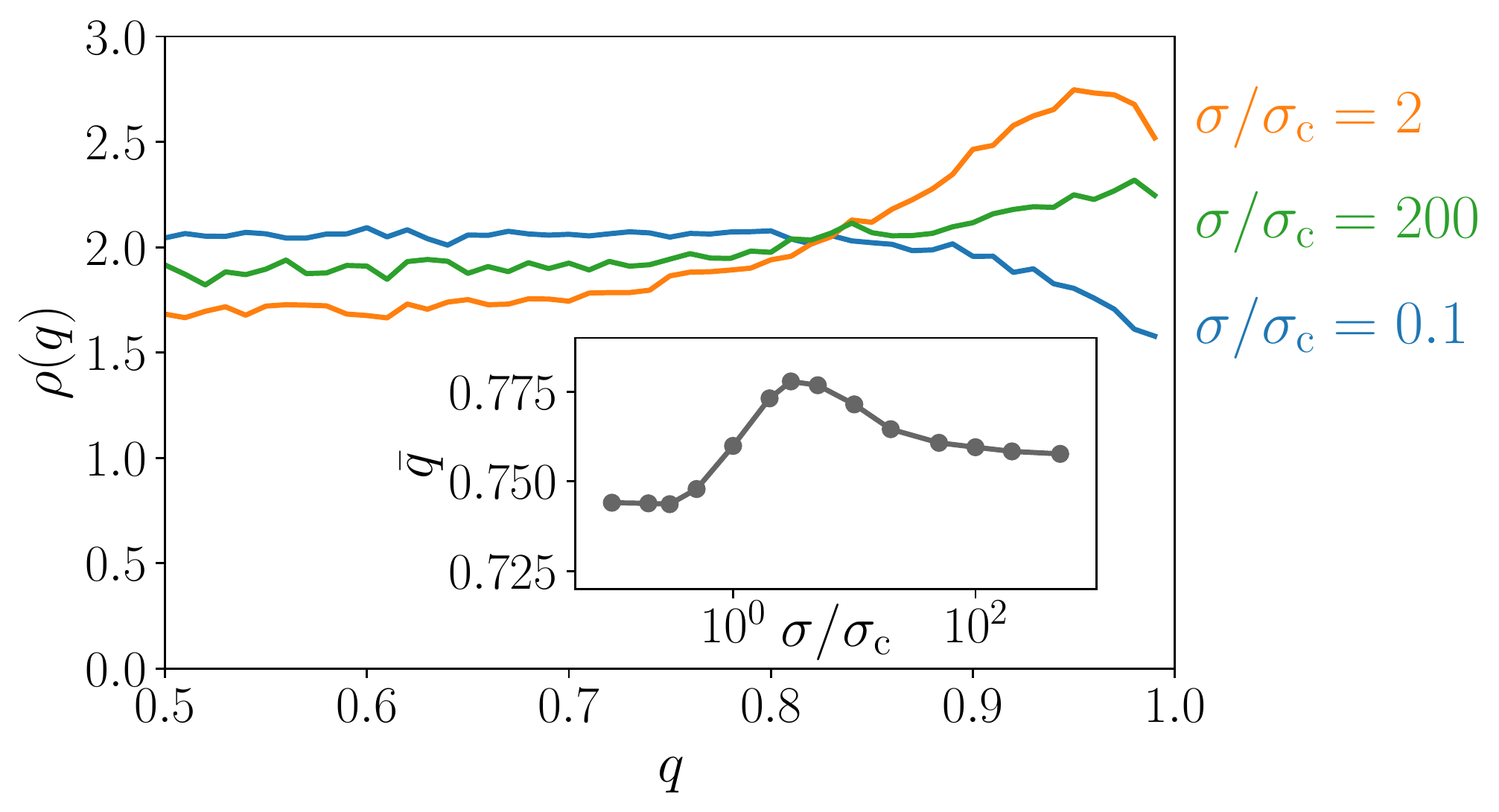}
  \caption{Distribution of \(q\) values measured in the critical-load model with sliding friction coefficient \(\mu_\mathrm{s}=1\), 
  and no rolling friction, at volume fraction \(\phi=0.5\) and three values of stresses 
  \(\sigma/\sigma_\mathrm{c}=0.1\) (below shear-thickening, in blue), 
  \(\sigma/\sigma_\mathrm{c}=2\) (during shear-thickening, in orange), 
  and \(\sigma/\sigma_\mathrm{c}=200\) (above shear-thickening, in green). 
  In inset, the average value \(\bar{q}\) as a function of the applied stress, for the same volume fraction.}\label{fig:q_sliding}
\end{figure}

Within the \(q\)-model perspective, this can be seen in the values of \(q\) measured in the simulations, 
across shear thickening.
To measure \(q\) in the CLM, we follow the procedure described in~\cite{coppersmith_model_1996}. 
The \(q\)-model describes the propagation of the force along the principal stress axis, 
which in our case is the compressional direction \(\hat{c} = \{-1/\sqrt{2}, 1/\sqrt{2}, 0 \} \).
For each particle, we decompose the contact force network 
into incoming forces projected on \(\hat{c}\) 
from contacts in a direction \(\hat{n}\) such that \(\hat{n}\cdot\hat{c}>0\) 
and outgoing forces projected on \(\hat{c}\) from contacts with \(\hat{n}\cdot\hat{c}<0\).
(Note that in our case, contrary to granular packings under gravity, there is no ``up'' or ``down'' directions, 
i.e., ``incoming'' and ``outgoing'' labels are purely conventional and interchangeable.)
Consistently with our specific setup of the \(q\)-model, 
we select particles with 2 outgoing contacts (a typical case for our simulations), 
and for these we compute the ratio between 
the largest outgoing force and the smallest one, that is, the ratio \(q/(1-q)\) for this site.
Because the system is disordered, each of these particles sees a different environment, and instead of 
having a single value of \(q\) for all particles like in Eq.~\ref{eq:Pofq_qmodel}, 
we have a distribution \(\rho(q)\).
Fig.~\ref{fig:q_sliding} shows 
the distributions \(\rho(q)\) obtained with different applied stresses.
We indeed see that in the thickened state (\(\sigma/\sigma_\mathrm{c}=200\)) the distribution of \(q\) drifts to larger values, 
indicative of a more heterogeneous transmission, compared to the unthickened state (\(\sigma/\sigma_\mathrm{c}=0.1\)). 
This is confirmed by the average value \(\bar{q}\) being larger above than below thickening (in inset). 
Interestingly, however, the trend across thickening is not monotonic: the largest \(\bar{q}\) can be seen 
during thickening, and this is confirmed by the full distribution \(\rho(q)\) at \(\sigma/\sigma_\mathrm{c}=2\) 
showing an accumulation of sites with large \(q\) values above 0.9. 
While this non-monotonic behavior is interesting (and indeed explains some aspects of the force distribution during thickening, 
as we will discuss later), we for now just assume that 
there are only 2 different values for \(q\):
a smaller one for the frictionless state and a larger one for the frictional state.

\begin{figure}[t]
    \centering
    \includegraphics[width=0.47\textwidth]{\figfolder/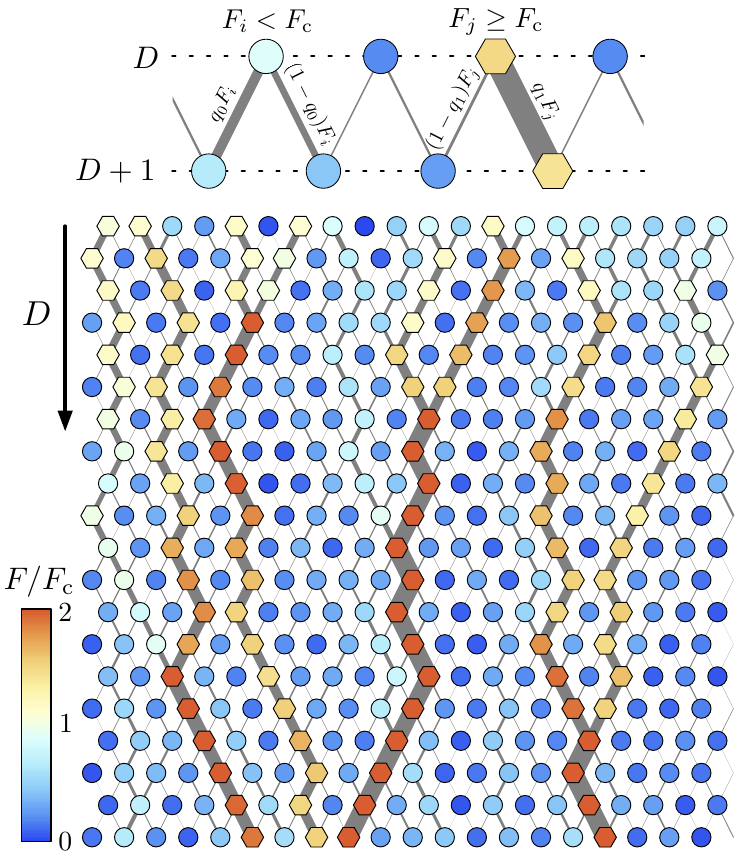}
    \caption{\textbf{Top:} Sketch of two successive layers of the bi-\(q\)-model. Sites on the top layer, \(D\), are connected to the two sites beneath them in layer \(D+1\). Forces are transmitted through these bonds, like in the usual \(q\)-model, with one neighbor receiving \(qF\) and the other one \((1-q)F\). However, the value of \(q\) depends on the value of \(F\): if \(F<F_\mathrm{C}\), \(q=q_0\), otherwise \(q=q_1>q_0\).
      \textbf{Bottom:}
      An example of force propagation in the bi-\(q\)-model.
      The force distribution
      of the average $\bar{F} = 0.7 F_{\mathrm{c}}$
      is set to the top layer, and the forces propagate to downward.
      Sites of $F < F_{\mathrm{c}}$ (circles) are frictionless and distribute 
      forces to two sites in the next layer with a smaller ratio \(q_0 = 0.82\),
      while sites of $F \geq F_{\mathrm{c}}$ (hexagons) are frictional 
      and distribute with \(q_1 = 0.92\).}\label{fig:biq_sketch}
\end{figure}

With this in mind, we then introduce a bi-\(q\)-model, which is a modification of the \(q\)-model.
In this model, the \(q\) value of a site in layer \(D\) 
depends on the force received from layer \(D-1\).
If the received force \(F\)  is such that \(F<F_\mathrm{c}\), 
the site is considered as ``frictionless''
and propagates the force by picking one site to receive \(q_0F\) and the other one \((1-q_0)F\).
If \(F>F_\mathrm{c}\), the site is considered as ``frictional''
and one neighbor receives a weight \(q_1F\) and the other one \((1-q_1)F\) (see Fig.~\ref{fig:biq_sketch}).
The cases we are interested in are such that \(q_1 > q_0\), that is, 
force transmission is more unfair when friction sets in.

We can readily expect that the bi-\(q\)-model will predict force distributions 
with fatter tails at large forces: 
whenever the force on a site exceeds the critical load, a larger share of this force will be transmitted to one neighbor, 
making this neighbor more likely to be frictional in turn and therefore more likely to transmit a large force to only one neighbor. 
At continuum level, diffusion of forces with depth is hindered in favor of random advection~\cite{claudin_models_1998}, 
leading to an accumulation of the load on linear force-bearing structures roughly along the depth direction, 
akin to the usual force chains of dry granular media~\cite{radjai_force_1996,behringer_predictability_1999}.

\begin{figure}[t]
    \centering
    \includegraphics[width=0.4\textwidth]{\figfolder/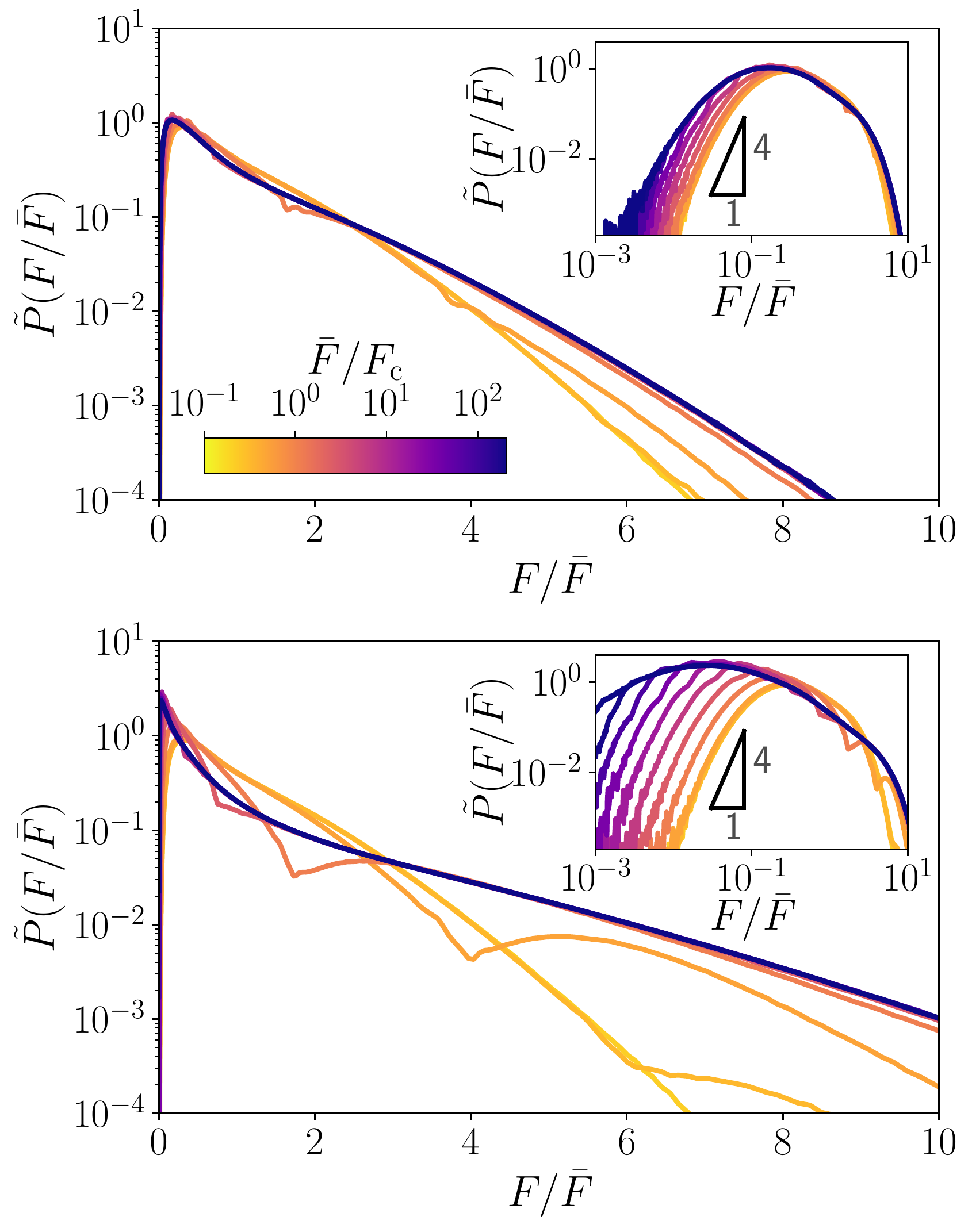}
    \includegraphics[width=0.4\textwidth]{\figfolder/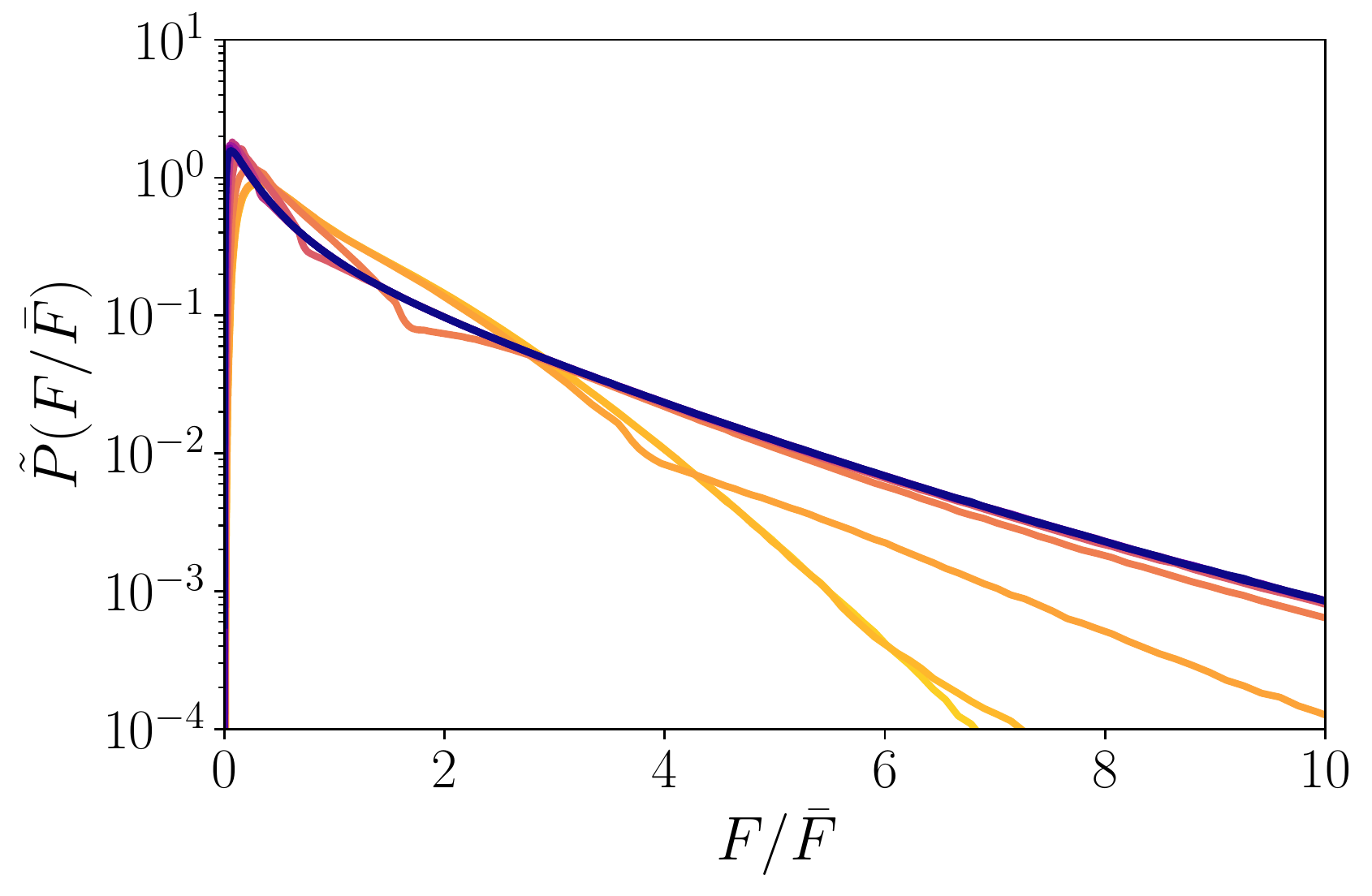}
    \caption{
    Steady-state force distributions \(\tilde{P}_\infty(F/F_\mathrm{c})\) 
      in the bi-\(q\)-model for \(q_0 = 0.82\) and
      \( q_1=0.86 \) (\textbf{top}) or \( q_1=0.92 \) (\textbf{middle}), in log-lin (main plots) and log-log scales (insets). 
    For each case, we show several values of \(\bar{F}/F_\mathrm{c}\) in order to cross the thickening transition, 
    from \(\bar{F}/F_\mathrm{c}=\times10^{-1}\) (yellow, below thickening) to \(\bar{F}/F_\mathrm{c}=2\times10^2\) 
    (dark blue, above thickening).
    \textbf{Bottom:} \(\tilde{P}_\infty(F/F_\mathrm{c})\) in the bi-\(q\)-model with \(q_0 = 0.82\) and \(q_1\) random
    with distribution uniform on \([0.82, 0.95] \).}\label{fig:PofF_biq}
\end{figure}

\begin{figure}[t]
    \centering
    \includegraphics[width=0.4\textwidth]{\figfolder/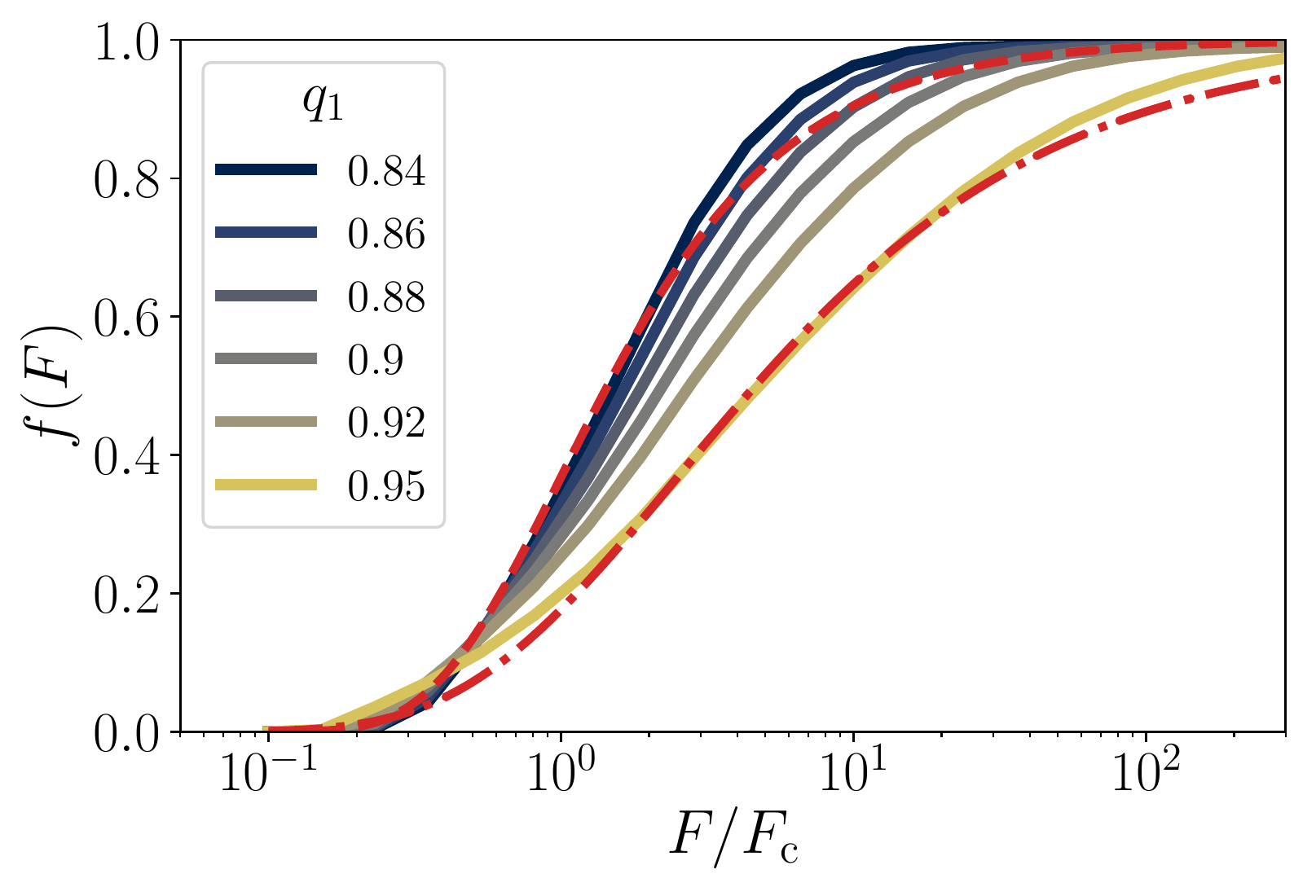}
    \caption{Fraction of frictional contacts \(f\) as a function of the applied average force \(\bar{F}/F_\mathrm{c})\) in the bi-\(q\)-model 
    for \(q_0=0.82\) and several values of \(q_1\). The red dashed line is \(\exp(-cF_\mathrm{c}/F)\) 
    with \(c=1\), and the red dashed-dotted line is \(\exp[-c'{(F_\mathrm{c}/F)}^\beta]\) with \(c' = 2.5\) and \(\beta=0.6\)}\label{fig:fofsigma_biq}
\end{figure}

We now investigate the behavior of the bi-\(q\)-model as a function of its parameter values \(q_0\) and \(q_1\).
We simulate a lattice with \(200\) layers of \(10^7\) sites.
We initiate the simulation by assigning forces on the first layer randomly picked, 
with a uniform distribution on \([0, 2\bar{F}]\) (such that the average force in each layer in \(\bar{F}\)). 
The force distribution is measured on the last layer only, and we checked that it does not evolve with the depth of the lattice 
any more by verifying that the distribution measured on a lattice of \(100\) layers is undistinguishable from the one we report.
The thickening transition is controlled in the bi-\(q\)-model by the parameter \(\bar{F}/F_\mathrm{c}\): 
when \(\bar{F}/F_\mathrm{c} \ll 1\), the system is in the low viscosity state, while when \(\bar{F}/F_\mathrm{c} \gg 1\), 
it is in the high viscosity state.

We pick \(q_0 = 0.82\) for the frictionless state, which gives force distributions at low \(\bar{F}/F_\mathrm{c}\) 
in qualitative agreement with what we observe in the frictionless state of particle-based simulations 
(yellow lines in Fig.~\ref{fig:PofF_biq}).
In particular, this value gives an exponential decay with a constant \(c\approx 1.9\), 
compatible with the one found in the particle-based simulations (see~Fig.~\ref{fig:PofF_simus_sliding}), 
although the exponential tail does not extend as low in forces in the model.
At lower values of \(q_0\) (below roughly \(0.7\)), 
we observe a faster decay \(\propto \exp\{-(F/\bar{F})^2\}\), 
in agreement with what is known for crystalline packings 
(which the \(q\)-model is when the force transmission is even)~\cite{tighe_force_2010}.

We show in the top panel of Fig.~\ref{fig:PofF_biq} the force distributions obtained for various values of \(q_1\) at \(q_0 = 0.82\). 
As in the particle-based simulations (Fig.~\ref{fig:PofF_simus_sliding}), 
for \(q_1 = 0.86\) we observe two limiting behaviors for small and large \(\bar{F}/F_\mathrm{c}\) values, 
corresponding to the distribution observed for a usual \(q\)-model with \(q=q_0$ (resp. \(q=q_1\)). 
They both show characteristic exponential tails, and the decay constant of the large \(\bar{F}/F_\mathrm{c}\) distribution 
is smaller then the one of the small \(\bar{F}/F_\mathrm{c}\) distribution.
At intermediate values of \(\bar{F}/F_\mathrm{c}\), the distributions tend to the small 
\(\bar{F}/F_\mathrm{c}\) distribution for \(F\ll F_\mathrm{c}\) 
and to the large \(\bar{F}/F_\mathrm{c}\) distribution for \(F \gg F_\mathrm{c}\), 
as if the system was separating into frictional and frictionless parts 
essentially similar to what would be observed in a purely frictional or frictionless system, respectively.
Note that we do not see here the behavior which was seen at intermediate stress values \(\sigma/\sigma_\mathrm{c}\) 
in the particle-based simulations, where the large force decay constant \(c\) was smaller than in the frictional state.
This is presumably an effect of the larger values of \(\bar{q}\) observed at these stresses, which we ignored in the 
bi-\(q\)-model, where \(q\) is not explicitly a function of the overall applied stress.
Also, the low-force end of the force distribution has a power law, like in the CLM simulations, 
but the exponent is much larger in the bi-\(q\)-model, we observe \(\theta\approx 4\).
The underestimation of the number of low forces is a known deficiency of the \(q\)-model~\cite{bo_cavity_2014}.

At \(q_1 = 0.92\), that is, when the contrast between force transmission in the unthickened and thickened states is large, 
the large-force tail in the thickened state is significantly altered and decays much slower than at lower \(q_1\). 
The most notable difference from the results of the CLM simulations and lower \(q_1\) bi-\(q\)-model is the existence of a minimum in 
\(\tilde{P}_\infty(F/F_\mathrm{c})\) around \(F/F_\mathrm{c} = 1\).
This is a direct effect of the brutal contrast in transmission when a site turns frictional. 
We can indeed get rid of this feature entirely by randomly picking \(q_1\) on each frictional site in a smoother distribution.
In the bottom of Fig.~\ref{fig:PofF_biq}, we show the case of an uniform distribution on an interval \([q_0, q^\mathrm{max}]\).
Nonetheless, the slower decay of the distribution in the thickened state remains.
This implies that the approximation of an exponential force distribution
which is independent of the applied stress, Eq.~\ref{eq:PofF_exp}, 
rapidly loses accuracy when \(q_1\) is far from \(q_0\).

We can then integrate the obtained force distributions to get the fraction of frictional contacts 
\(f(\bar{F})\) as function of the average force \(\bar{F}\) predicted by the bi-\(q\)-model, as shown in Fig.~\ref{fig:fofsigma_biq}. 
While for \((q_0, q_1)=(0.82,0.84)\) and \((0.82,0.86)\) (low contrast between force transmission in the unthickened and thickened states), 
the behavior is fairly consistent with Eq.~\ref{eq:exp_fofsigma}, unsurprisingly for largely different \(q_0\) and \(q_1\) values 
\(f(\bar{F})\) departs significantly from the exponential behavior.
Actually, for large values of \(q_1\) the predicted \(f(\bar{F})\) is better fitted 
by a stretched exponential \(\exp\bigl[-c{(F_\mathrm{c}/\bar{F})}^\beta\bigr]\),
with \(\beta\approx 0.6\) for \(q_1=0.95\).
This behavior is also seen when we pick \(q_1\) within a uniform distribution (not shown).

Going back to actual suspensions, the main prediction of the bi-\(q\)-model is then that, 
if the force transmission is much more uneven 
in the thickened state than in the thickened state, one should observe a much broader \(f(\sigma)\), 
which in turn implies thickening on a much wider stress range.
We will now test this idea in particle-based simulations.


\subsection{Rolling friction}

We know that, for suspensions of spherical particles with sliding friction, 
the force transmission contrast is not enough to depart from an exponential \(f(\sigma)\).
We can correlate the contrast between \(q_0\) and \(q_1\) to the difference between 
the contact numbers at frictionless jamming \(z_0\) and frictional jamming \(z_1\). 
We decide to play on \(z_1\), and simulate particles with rolling friction as well as sliding friction, which is known to affect \(z_1\) 
(and also the jamming point \(\phi_\mathrm{J}^1\)) significantly~\cite{zhou_rolling_1999,magalhaes_jamming_2014}, 
by restricting degrees of freedom more then sliding friction alone.
A Maxwell-like counting argument~\cite{maxwell_calculation_1864} indeed predicts 
that for the large sliding friction coefficient \(\mu_\mathrm{s}\) limit, 
\( z^1_0 = d+1 \) (with \(d\) the spatial dimension of the system, \(z^1_0=4\) for our simulations with \(d=3\)) 
without rolling friction, and \(z^1_0 = d(d+1)/(2d-1)\) (\(z^1_0=12/5\) in \(d=3\)) 
in the large rolling friction coefficient \(\mu_\mathrm{r}\) limit.

\begin{figure}[t]
    \centering
 \includegraphics[width=0.4\textwidth]{\figfolder/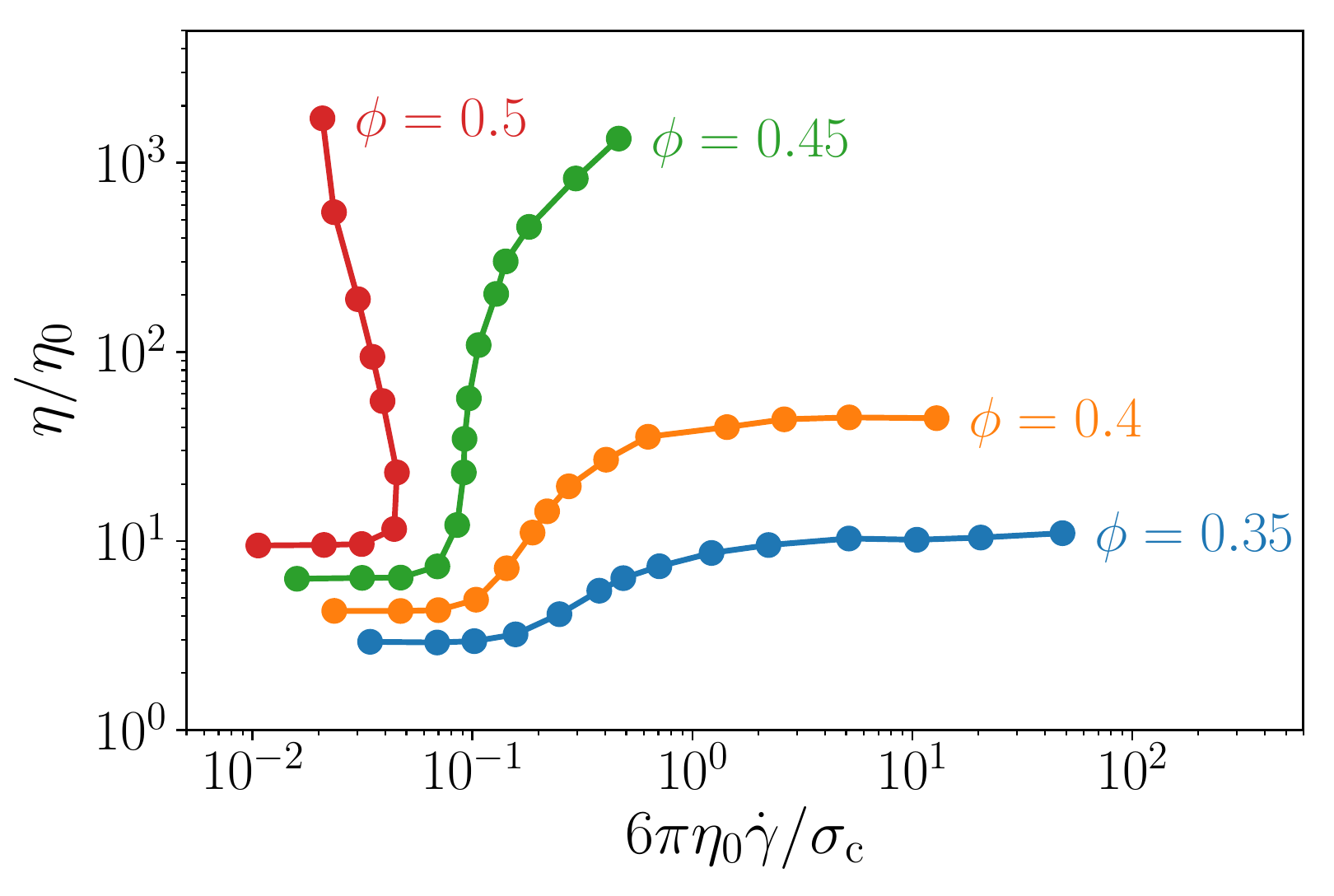}
 \includegraphics[width=0.4\textwidth]{\figfolder/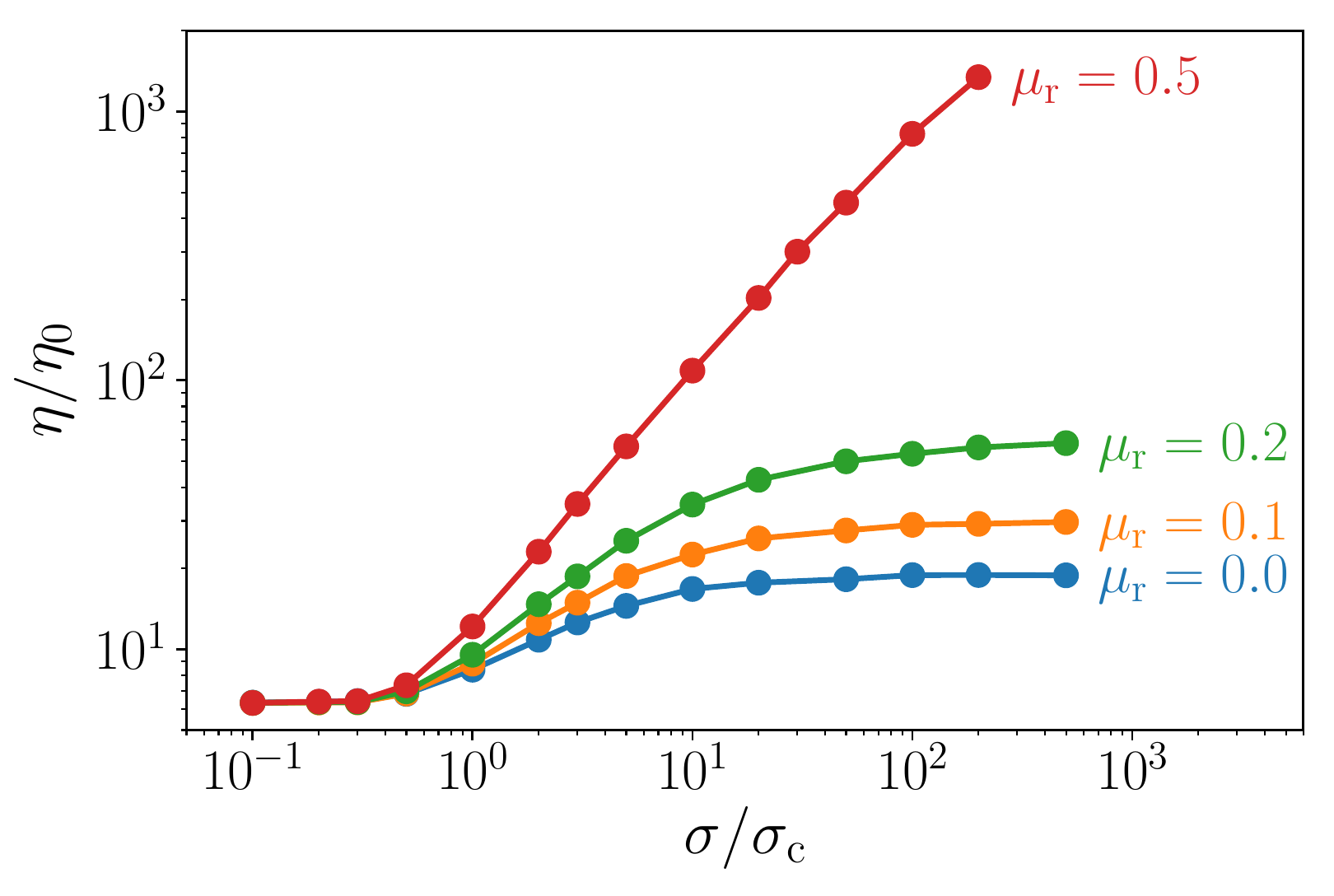}
    \caption{\textbf{Top:} Relative viscosity as a function of shear rate for the CLM with sliding friction coefficients \(\mu_\mathrm{s}=1\) and  
    rolling friction coefficient \(\mu_\mathrm{r}=0.5\), for several volume fractions. 
    \textbf{Bottom:} Relative viscosity as a function of shear stress for sliding friction coefficients \(\mu_\mathrm{s}=1\) and  
    rolling friction coefficients \(\mu_\mathrm{r}=0,\ 0.1,\ 0.2\) and \(\mu_\mathrm{r}=0.5\), at a fixed \(\phi=0.45\).}\label{fig:rheo}
\end{figure}

We keep the sliding friction coefficient \(\mu_\mathrm{s} = 1\), 
and study the behavior under imposed shear stress, varying \(\mu_\mathrm{r}\).
First, we show that the shear-thickening rheology is qualitatively unchanged by the addition of rolling friction, 
as seen in Fig.~\ref{fig:rheo}, 
with continuous shear thickening (CST) at low volume fractions turning to a discontinuous shear thickening (DST) at high volume fractions.
Quantitatively, however, this phenomenology is entirely shifted towards lower volume fractions, because jamming for grains with both sliding 
and rolling friction happens at significantly lower volume fraction 
than the jamming point for systems with sliding friction only. 
For instance, DST appears around \(\phi=0.45\) for \(\mu_\mathrm{r} = 0.5\), 
but it appears around \(\phi=0.56\) for \(\mu_\mathrm{r} = 0\)~\cite{Mari_2014,mari_nonmonotonic_2015}.

\begin{figure}[t]
    \centering
 \includegraphics[width=0.4\textwidth]{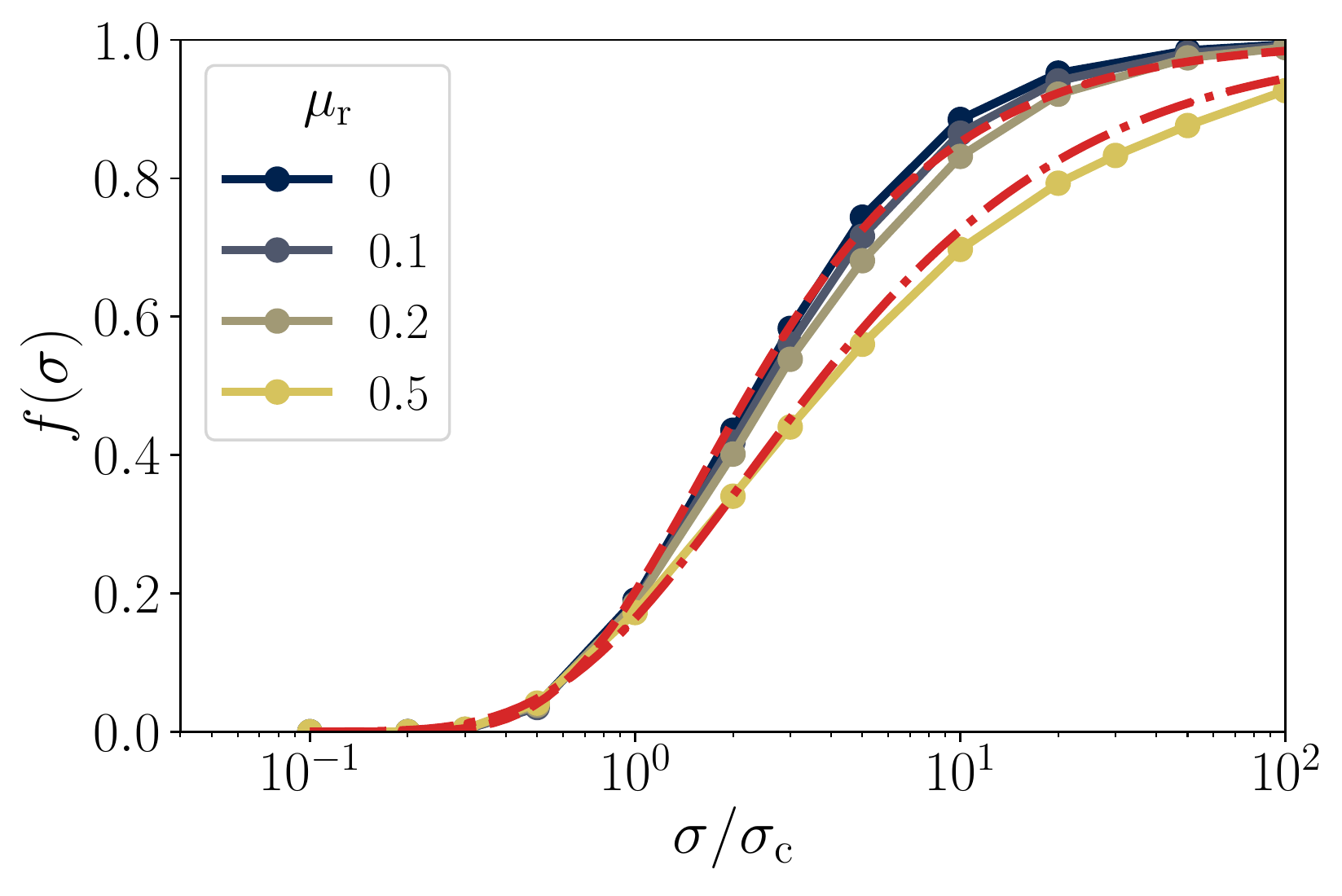}
 \includegraphics[width=0.4\textwidth]{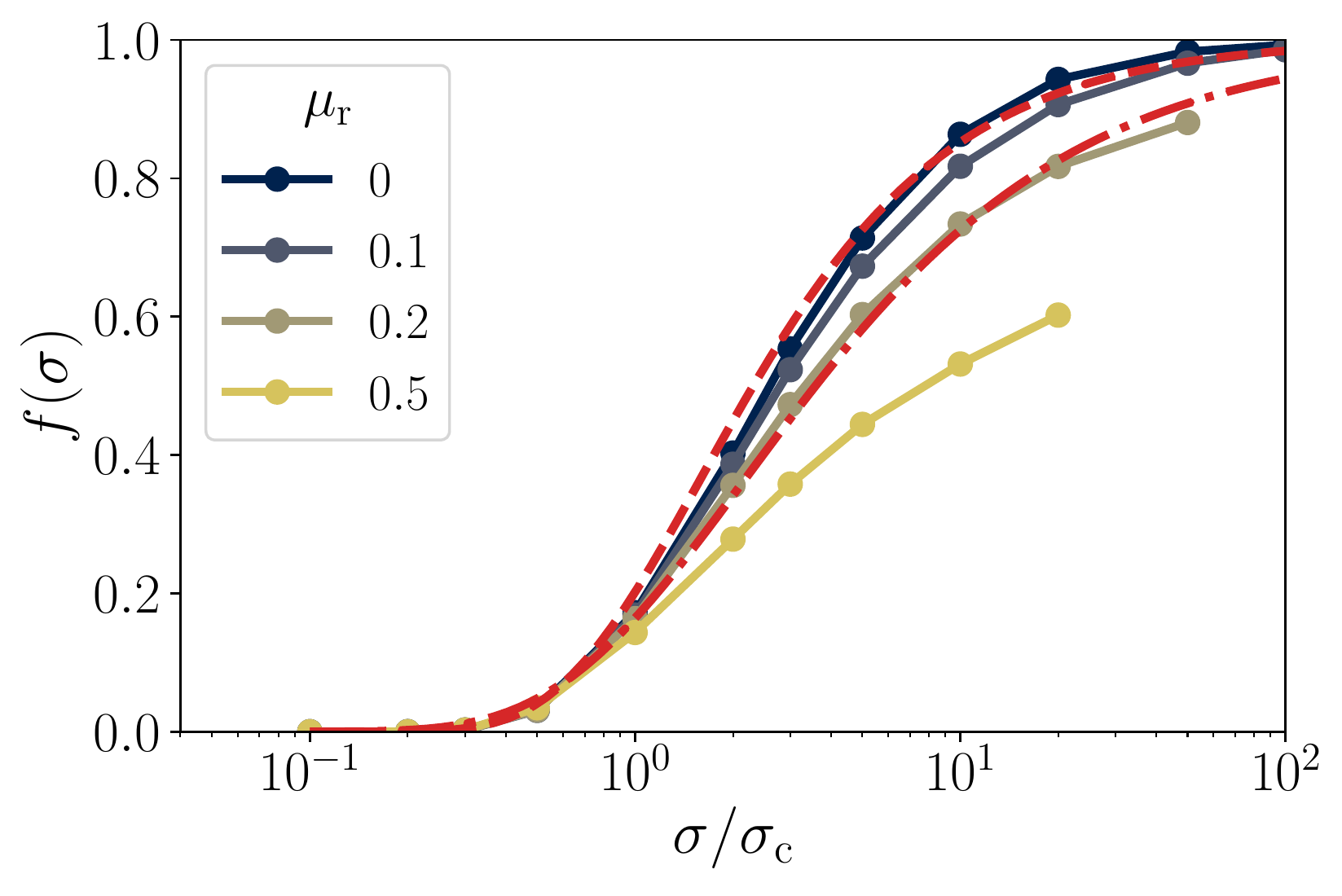}
    \caption{Fraction of frictional contacts \(f\) as a function of applied stress \(\sigma/\sigma_\mathrm{c}\) in the CLM with 
    \(\mu_\mathrm{s}=1\) and several rolling friction coefficients \(\mu_\mathrm{r}\), at volume fractions \(\phi=0.45\) (\textbf{top})
    and \(\phi=0.5\) (\textbf{bottom}).
    }\label{fig:fofsigma_simus}
\end{figure}

Another observation is that shear thickening takes place on a rapidly growing range of stresses when \(\mu_\mathrm{r}\) increases.
While thickening starts at an onset stress \(\sigma_{\mathrm{on}}\) independent of \(\mu_\mathrm{r}\), 
thickening stops at \(\sigma/\sigma_{\mathrm{on}}\approx 10^2\) for \(\mu_\mathrm{r}=0\) 
but only at \(\sigma/\sigma_{\mathrm{on}}\approx 10^3\) for \(\mu_\mathrm{r}=0.2\).
This feature is also present in the fraction of frictional contacts \(f(\sigma)\), shown in Fig.~\ref{fig:fofsigma_simus} 
for \(\phi=0.45\) and \(\phi=0.5\). It appears that the sigmoidal behavior of \(f(\sigma)\) 
also happens on a significantly wider range of stresses for \(\mu_\mathrm{r}=0.5\) than for \(\mu_\mathrm{r}=0\).
Actually, the \(\mu_\mathrm{r}=0.5\) data strongly deviate from a \(\exp(-\sigma^\ast/\sigma)\) curve, 
which we can fit better for \(\phi=0.45\) with a stretched form 
\(\exp\bigl(-{(\sigma^\ast/\sigma)}^b\bigr)\), with \(b=0.75\). 
This is consistent with the predictions of the bi-\(q\)-model presented in the previous section.
However, while the fit is acceptable for \(\phi=0.45\), it is completely off for \(\phi=0.5\).
Indeed, the \(f(\sigma)\) relation is no more insensitive to volume fraction changes, as it was for the \(\mu_\mathrm{r}=0\) case.
From the point of view of the bi-\(q\)-model, these results are consistent with an increase of \(q_1\) when \(\phi\) increases.

\begin{figure}[t]
    \centering
    \includegraphics[width=0.38\textwidth]{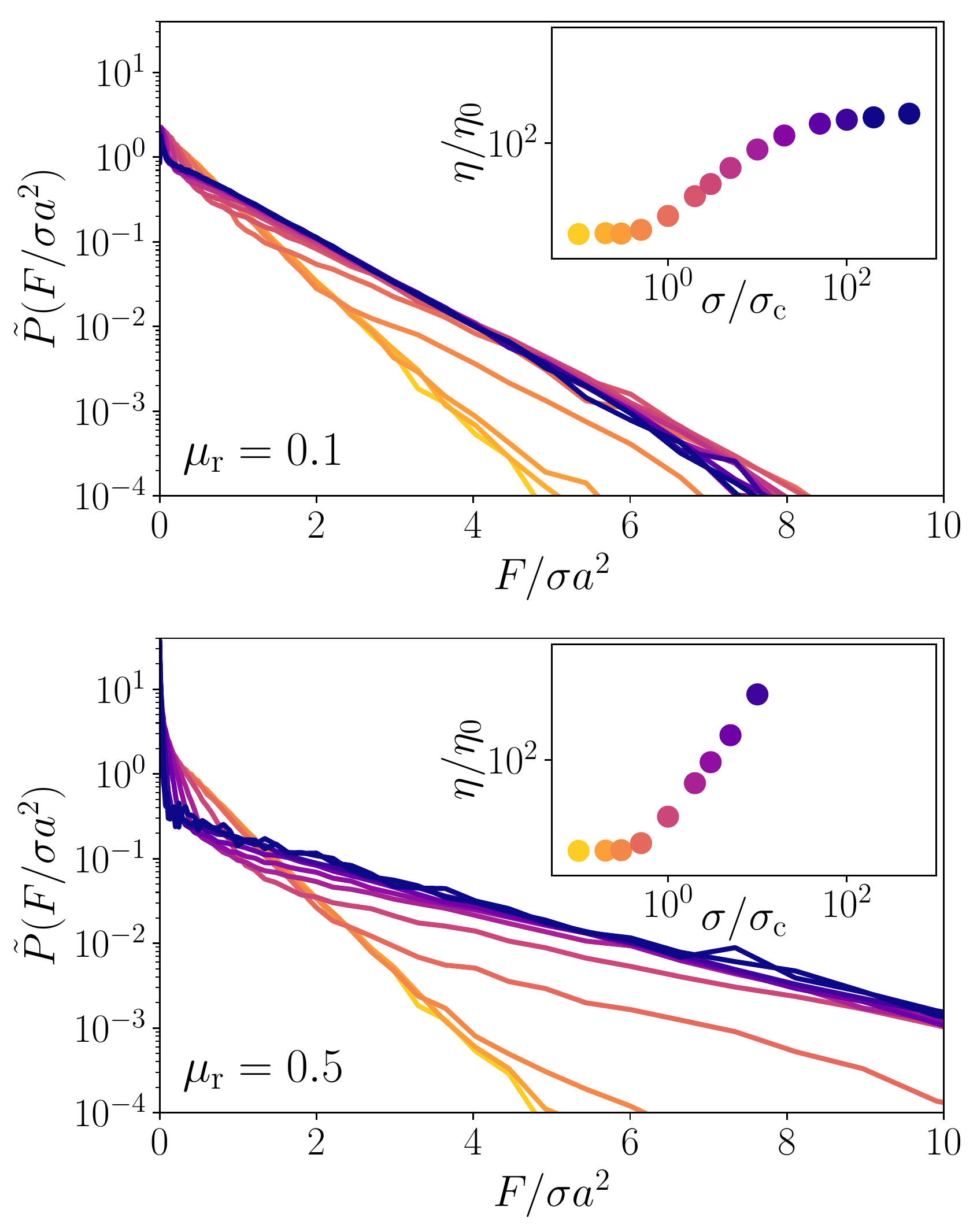}
    \caption{Force distribution \(P(F)\) in simulations of the CLM model for sliding friction coefficient \(\mu = 1\) and rolling friction coefficients 
    \(\mu_\mathrm{r}=0.1\) (top) and \(\mu_\mathrm{r}=0.5\) (bottom). 
    For each plot, the several curves correspond to different values of applied stress \(\sigma/\sigma_0\) across the shear-thickening transition. 
    Each curve is colored accordingly to the symbols in the viscosity vs.\ stress flow curves shown in the insets.}\label{fig:PofF_simus_rolling}
\end{figure}
In turn, these observations have their counterpart at the microscopic scale in the distribution of normal contact forces, 
which confirms the predictions of the bi-\(q\)-model. 
We show this distribution in Fig.~\ref{fig:PofF_simus_rolling} for \(\mu_\mathrm{r}=0.1\) and \(\mu_\mathrm{r}=0.5\), both at \(\phi=0.5\).
As for the pure sliding friction case, the two limiting behaviors at low and high stresses are mostly exponential \(P(F)\approx \exp(-cF)\).
At low stresses, we again find \(c_0\approx 2\), as it is the same frictionless state than for the simulations shown in Fig.~\ref{fig:PofF_simus_sliding}.
At larges stresses, we find \(c_1 \approx 1.1\) (resp. \(c_1 \approx 0.5\)) for \(\mu_\mathrm{r}=0.1\) (resp. \(\mu_\mathrm{r}=0.5\)).
This in particular implies that for a given applied stress \(\sigma \), the number of very large forces (say \(F/\sigma a^2>10\)) 
becomes significant when rolling friction sets in, 
when it was essentially absent for system with only sliding friction, 
which is exactly what the bi-\(q\)-model predicts (see Fig.~\ref{fig:PofF_biq}).


\section{Conclusion}

We introduced an extension of the celebrated \(q\)-model (which we call bi-\(q\)-model) 
of force propagation in granular matter~\cite{liu_force_1995,coppersmith_model_1996}, 
intended at describing the force distribution observed in shear-thickening suspensions, 
which is itself directly related to the rheology through the Wyart-Cates model~\cite{Wyart_2014}.
It provides a tool to simply evaluate, at least qualitatively, the effect of microscopic interaction details on the global rheology.
It rationalizes the fact that for the model case of spherical particles with sliding friction 
and fairly monodisperse size distribution, the non-hydrodynamics force distribution does not evolve much during thickening 
and stays close to a distribution with an exponentially decaying tail, which was argued to be the source of the peculiar 
exponential relation between the fraction of frictional contacts \(f\) and stress \(\sigma \) 
(Eq.~\ref{eq:exp_fofsigma}) observed in these systems.

It highlights that a central aspect of the force distribution is the evenness with which a particle typically distributes its load on its neighbors.
Indeed, it predicts that if this force ``diffusion'' is largely different between the low viscosity and the high viscosity 
states, the force distribution during thickening should strongly deviate from the usual exponential tail at large forces, 
and in consequence shear thickening will occur on a much broader stress range.
We showed that this is well verified in particle-based simulations: when the high viscosity state is dominated by rolling friction 
(and not only sliding friction), the force distribution shows a wide range of non-exponential decay at large forces during thickening, 
and the subsequent \(f(\sigma)\) is a much broader sigmoidal function than Eq.~\ref{eq:exp_fofsigma}.

We do not however expect quantitative agreement with actual systems from a simple model like the bi-\(q\)-model.
Many variations on the same lines of thought could be developed in order to achieve a seemingly better fit with simulation data 
(like we did in the bottom of Fig.~\ref{fig:PofF_biq}), 
but we feel it would be (at this stage at least) 
a fine tuning from which not much can be learned regarding the physics of shear thickening.
The main point of our work is to show that, with simple physical considerations, one can adapt a force propagation model
and extract the correct qualitative effect of changes at the level of particle contacts on the macroscopic rheology, 
via the stress dependent fraction of frictional contacts \(f(\sigma)\) appearing in the Wyart-Cates model.

\begin{figure}[t]
    \centering
    \includegraphics[width=0.49\textwidth, trim=0 -10 0 0]{\figfolder/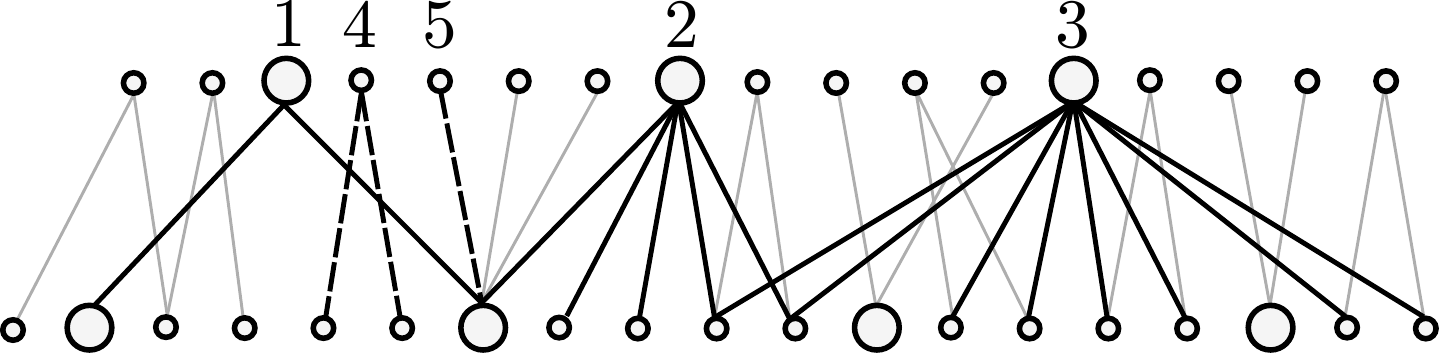}
    \includegraphics[width=0.4\textwidth]{\figfolder/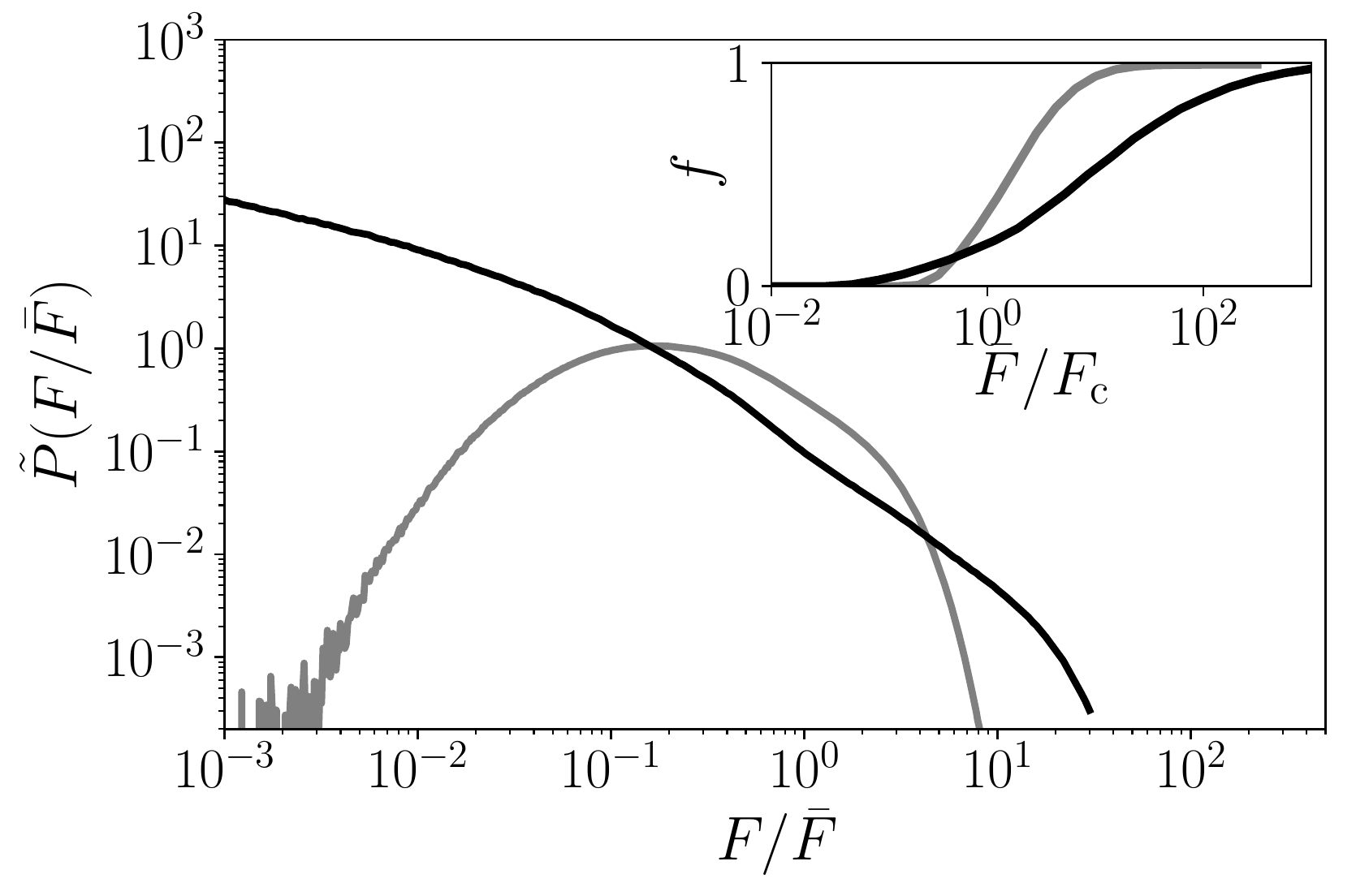}
    \caption{Inclusion of bidispersity in the bi-\(q\)-model. \textbf{Top:} Sketch of the different bond cases between big and small sites. Big sites (labeled 1, 2 and 3) can propagate their forces to either 2 big sites (site 1), 1 big/\(k\) small (site 2), or \(2k\) small sites (site 3), each case being picked randomly with equal probabilities. Small sites can propagate to 2 small sites (site 4) or 1 big site (site 5). 
    \textbf{Bottom:} Typical force distribution from the bidisperse bi-\(q\)-model with \(q_0=0.82\) and \(q_1=0.86\), in black, alongside the force distribution from the usual bi-\(q\)-model at the same average force and \(q_0, q_1\), in grey. 
    In inset, predictions for the fraction of frictional contacts \(f(\sigma)\), for the usual (in grey) and bi-disperse (in black) bi-\(q\)-models.}\label{fig:biq_bidisperse}
\end{figure}

While we considered here only the effect of rolling friction, through the introduction of a stress sensitive \(q\)-distribution, one 
can study at the same qualitative level other modifications of the canonical suspension of spheres with sliding friction,
on which many experimental and numerical works so far focused, 
in order to approach the much broader diversity of real-world suspensions~\cite{guy_constraint-based_2018}.
For instance, a large bidispersity has also recently been argued to broaden the sigmoidal shape of \(f(\sigma)\)~\cite{guy_testing_2019}. 
We can model this point by a simple modification of our bi-\(q\)-model, to include ``big'' and ``small'' sites (see top panel of Fig.~\ref{fig:biq_bidisperse}).
The main microscopic effect of the large size ratio between small and large particles is steric: around any given particle, 
a contact with a big particle will occupy a large solid angle excluded to any other particles.
We then modify the bi-\(q\)-model such that every \(n^\mathrm{th}\) site is ``big'', and 
can transmit its force with equal probability to either (i) 2 other big sites beneath it, (ii) 1 big and \(k>1\) small sites, or (iii) \(2k\) small sites.
Similarly, small sites can propagate only to 2 small sites or 1 big.
If a site has only one downward bond, it gives the entirety of its force \(F\) to this downward neighbor. 
If it has \(m\geq 1\) downward neighbors, it gives \(qF\) to one of them, and \((1-q)F/(m-1)\) to the \(m-1\) others.
In the bottom panel of Fig.~\ref{fig:biq_bidisperse}, we show the prediction of this bidisperse bi-\(q\)-model regarding the force distribution 
with \(k=4\) and \(n=10\), and values of \(q\) used to mimic shear thickening of particles with sliding friction only, 
\(q_0=0.82\) and \(q_1=0.86\).
It has no exponential tail at large forces, nor maximum at low forces, and is markedly different from the monodisperse bi-\(q\)-model prediction, 
also shown in the same figure.
These features are strikingly similar to the normal force distribution obtained from simulations 
of highly polydisperse dry granular packings~\cite{voivret_multiscale_2009}.
Remarkably, we found it is almost insensitive to the average force, i.e., the force distribution is not predicted to evolve much across thickening.
Finally, the subsequent \(f(\bar{F})\) is indeed predicted to be much broader than~Eq.~\ref{eq:exp_fofsigma} as is seen in simulations~\cite{guy_testing_2019}, 
and cannot either be well fitted by a stretched exponential 
\(\exp\bigl[-c{(F_\mathrm{c}/\bar{F})}^\beta \bigr]\).

As discussed in the introduction, the original \(q\)-model was criticized for having the wrong continuum limit~\cite{claudin_models_1998}: 
at large scales, the stress \emph{diffuses}, which is not what is observed in granular packings, and cannot explain, for instance, 
the well-known pressure minimum observed beneath the apex of a sandpile~\cite{jotaki_bottom_1979,vsmid1981pressure}.
One could then be genuinely worried about the validity of the model concerning the force distribution.
Remarkably, however, as far as the force distribution is concerned, the \(q\)-model behaves very much like a mean-field model, 
and indeed the mean-field solution for the force distribution is known to be exact for a specific (uniform) 
distribution of \(q\)~\cite{coppersmith_model_1996}.
This quasi mean-field nature implies that spatial aspects of stress propagation at large scales are irrelevant for the force distribution.
We could verify this in the bi-\(q\)-model too, with a mean-field version of the model for which each site does not transmit its force 
to the 2 neighbor beneath it but to 2 neighbors randomly picked within the layer below: 
this gives force distributions virtually undistinguishable from the ones obtained with the lattice version of the bi-\(q\)-model, except for 
a slightly smaller exponent \(\theta \) at low forces.

Extensions of the \(q\)-model could then be used as a design tool for shear-thickening suspensions with a taylored, desired thickening behavior.
Many aspects of the link between microscopics and steady-state rheology of shear-thickening suspensions remain to be explored and understood however.
Other parts of the Wyart-Cates model may be fragile with respect to changes in the microscopic details, 
in particular the relation between jamming point and fraction of frictional contact~\cite{guy_testing_2019}, 
and relation between viscosity and distance to jamming point (recent results suggest a different divergence exponent 
for non-thickening suspensions of rods than for suspensions of spheres~\cite{tapia_rheology_2017}).
Nonetheless, the fact that a mean-field model like WC, as well as the quasi mean-field extensions 
to the \(q\)-model can correctly predict many non-trivial rheological trends of thickening suspensions 
leaves the possibility that quantitative predictive tools could just come from minute modifications from these simple models, 
as opposed to order-of-magnitude more complex many-body descriptions.

\section*{References}

%

\end{document}